\documentclass[final,3p,times,twocolumn]{elsarticle}

\usepackage{amssymb}
\usepackage{amsmath}
\usepackage{array}

\usepackage{amsfonts}
\usepackage{algorithmic}
\usepackage{algorithm}
\usepackage{textcomp}
\usepackage{stfloats}
\usepackage{url}
\usepackage{verbatim}
\usepackage{graphicx}
\usepackage{multirow}
\usepackage{color}

\journal{Speech Communication}

\begin{document}

\begin{frontmatter}



\title{Modeling and Estimation of Vocal Tract and Glottal Source Parameters Using ARMAX-LF Model}


\author[label1]{Kai Li}
\affiliation[label1]{organization={Graduate School of Advanced Science and Technology},
            addressline={Japan Advanced Institute of Science and Technology}, 
            city={Nomi},
            postcode={923-1292}, 
            state={Ishikawa},
            country={Japan}}

\author[label1]{Masato Akagi}

\author[label2]{Yongwei Li}
\affiliation[label2]{organization={National Laboratory of Pattern Recognition},
            addressline={Institute of Automation}, 
            city={Chinese Academy of Sciences},
            postcode={100101}, 
            state={Beijing},
            country={China}}

\author[label1]{Masashi Unoki}

\begin{abstract}
Modeling and estimation of the vocal tract and glottal source parameters of vowels from raw speech can be typically done by using the Auto-Regressive with eXogenous input (ARX) model and Liljencrants-Fant (LF) model with an iteration-based estimation approach. However, the all-pole autoregressive model in the modeling of vocal tract filters cannot provide the locations of anti-formants (zeros), which increases the estimation errors in certain classes of speech sounds, such as nasal, fricative, and stop consonants. In this paper, we propose the Auto-Regressive Moving Average eXogenous with LF (ARMAX-LF) model to extend the ARX-LF model to a wider variety of speech sounds, including vowels and nasalized consonants. The LF model represents the glottal source derivative as a parametrized time-domain model, and the ARMAX model represents the vocal tract as a pole-zero filter with an additional exogenous LF excitation as input. To estimate multiple parameters with fewer errors, we first utilize the powerful nonlinear fitting ability of deep neural networks (DNNs) to build a mapping from extracted glottal source derivatives or speech waveforms to corresponding LF parameters. Then, glottal source and vocal tract parameters can be estimated with fewer estimation errors and without any iterations as in the analysis-by-synthesis strategy. Experimental results with synthesized speech using the linear source-filter model, synthesized speech using the physical model, and real speech signals showed that the proposed ARMAX-LF model with a DNN-based estimation method can estimate the parameters of both vowels and nasalized sounds with fewer errors and estimation time.

\end{abstract}




\end{frontmatter}


\section{Introduction}
\label{sec:introduction}
Estimating glottal source and vocal tract parameters from speech signals using a fully parameterized glottal source and vocal tract model is crucial in many research fields, such as speaker verification~\cite{enzinger2011speaker, enzinger2011logarithmic}, emotional speech recognition~\cite{li2017commonalities,li2018contributions}, parametric speech synthesis~\cite{perrotin2020glottal,juvela2019glotnet,keller1995fundamentals}, and for further understanding of speech production mechanisms. The source-filter assumption, which models speech signals on the basis of exciting a vocal tract filter with a glottal source signal, is one of the most common assumptions for speech production processes and is extensively applied to speech analysis and synthesis \cite{wang2019neural,rao2018psfm,titze2016sensitivity,rao2018glottal}. 

Numerous methods of estimating glottal source waveforms and vocal tract shapes have been developed on the basis of the source-filter assumption of speech production. Linear prediction (LP) analysis is a traditional speech analysis method~\cite{atal1971speech, makhoul1975linear,makhoul1973spectral} which models the vocal tract and lip radiation in the same linear autoregressive (AR) model while assuming white noise excitation as input. The drawback of this method is that the estimated spectral envelope describes the characteristic of the vocal tract transfer function and contains information about the glottal source waveform.



\begin{figure*}[t]
\begin{center}
\includegraphics[width=140mm]{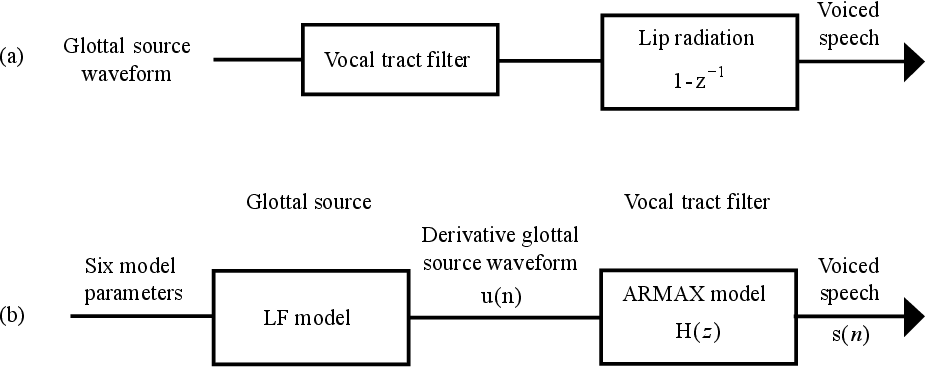}
\caption{Source-filter model (a) and its simplification using ARMAX-LF model (b) for voiced speech.}
\end{center}
\label{fig:SF}
\end{figure*}


To obtain the approximate value of a vocal tract filter, LP analysis~\cite{makhoul1975linear} is typically used by incorporating a glottal source model~\cite{lee1992robust, rabiner1978digital}, which is simply assumed as a series of pulses for voiced speech or white noise signals for unvoiced speech. These methods, however, are based on a simplistic assumption in the glottal source model, which cannot separate the characteristics of the glottal source waveform and vocal tract filter.

More elaborate models have been proposed to approximate the glottal source waveform or its derivatives, such as the Liljencrants-Fant (LF)~\cite{fant1985four}, Fujisaki-Ljungqvist (FL)~\cite{fujisaki1986proposal}, and Rosenberg-Klatt (RK) models~\cite{klatt1990analysis}. These sophisticated glottal source models combined with the autoregressive exogenous (ARX) model~\cite{li2017commonalities, ding1995simultaneous, fujisaki1996estimation, vincent2005estimation} on the basis of a joint-optimization process further improved the estimation accuracy. For example, Li et al.~\cite{li2020simultaneous,li2021f_0, takahashi2018estimation} estimated the parameters of the glottal source model and the vocal tract filter simultaneously based on the ARX model with the LF model (ARX-LF model) using an iterative algorithm under an analysis-by-synthesis methodology. However, the all-pole AR model in vocal tract modeling cannot provide anti-formant (zero) positions that frequently appear in certain classes of speech sounds, such as nasal, fricative, and stop consonants \cite{rahman2007identification}.

Theoretically, the linear predictive autoregressive moving average (ARMA) model estimates the vocal tract transfer function more reasonably than that based on an AR model. Formants (poles) and anti-formants (zeros) are calculated from the roots of the denominator and numerator polynomial given by the ARMA transfer function. There are many zeros in nasal, fricative, and stop consonants ~\cite{ouaaline1998pole,morikawa1982adaptive}. Zeros can suppress the peaks and flatten the spectrum in the frequency domain of a vocal tract filter. Achieving accurate estimation in this type of speech by a finite number of poles in the all-pole AR model is not easy. With the ARMA model, pole-zero characteristics in the vocal-tract transfer function are assumed. The ARMA model can provide information on zeros by using a low-order estimation~\cite{cadzow1980high}.

Conventional methods for ARMA model estimation~\cite{kopec1977speech,morikawa1982adaptive} generally operate on an estimate of raw speech, thus ignoring the complicated spectral characteristics of the glottal source waveform. As a consequence, formants and anti-formants concealed in the glottal source waveform could be mistakenly interpreted as vocal-tract formants and anti-formants, and the estimated formants and anti-formants are related to the spectrum of input speech, not a physiological vocal tract. In~\cite{fujisaki1996estimation,rahman2007identification}, the glottal source and the vocal tract are represented by separate models. However, it is still difficult to optimize multiple parameters.

We aim to extend the ARX-LF model to a wider variety of speech sounds. 
Toward this end, we previously proposed an autoregressive moving average exogenous (ARMAX)-LF model, or ARMAX-LF model, to represent the physiological processes of speech production~\cite{li2021study}. The ARMAX-LF model approximately represents the vocal tract as a pole-zero filter with an additional exogenous input (LF excitation) as an ARMAX model and derivative of the glottal source waveform as an LF model. The ARMAX model is much more effective than the conventional ARMA model because the introduction of additional exogenous LF excitation in the modeling of the vocal tract filter can decompose the source and filter from raw speech, hence decreasing the effects of the glottal source properties when estimating the vocal tract filter. However, two issues remain in the previous estimation procedure: (\textrm{i}) The accuracy of LF model initialization is low, which decrease the performance of simultaneous estimation using an iteration-based approach, especially in nasalized sounds. (\textrm{ii}) The iterative method requires estimating and updating glottal source and vocal tract parameters in every iteration, which makes the methods very expensive in computation. 

To overcome these remaining problems, in this paper, we first utilize the powerful nonlinear fitting capability of deep neural networks (DNNs) instead of mathematical methods to build a mapping from extracted acoustic speech features or speech waveforms to corresponding LF parameters. Then, accurate glottal source and vocal tract parameters are estimated without any iterations. The results for a large number of synthesized speech using source-filter and physical models and natural speech indicate that the ARMAX-LF model combined with the proposed estimation procedure can not only estimate glottal source and vocal tract parameters with fewer errors for both vowels and consonants but also reduce the estimation time significantly.

\section{Source-filter model of speech production}
As shown in Fig.~\ref{fig:SF}(a), the glottal source waveform, vocal tract, and lip radiation are represented linearly and non-interactively in the linear source-filter model of speech production. According to the source-filter assumption, the vocal tract and lip radiation filters are commutative and the effects of these filters can be represented by the derivative of the glottal source waveform. Thus, we can obtain a simplified form of the source-filter model, as shown in Fig.~\ref{fig:SF}(b).

Numerous methods for modeling the speech production process are based on the source-filter model. Among source-filter models, the ARX-LF model is frequently selected to represent the glottal source waveform and the vocal tract shape~\cite{vincent2005estimation, li2020simultaneous,agiomyrgiannakis2009arx} due to its overall adaptability to common speech waveforms and flexibility for representing extreme phonations~\cite{fant1985four,fu2006robust}. In this section, the ARX-LF model is introduced first, followed by the proposed ARMAX-LF model.

\subsection{ARX-LF model}

The ARX-LF model represents the derivative of the glottal source waveform by using the LF model and vocal tract filter by using an ARX model.

\begin{figure}[t]
\begin{center}
\includegraphics[width=0.47\textwidth]{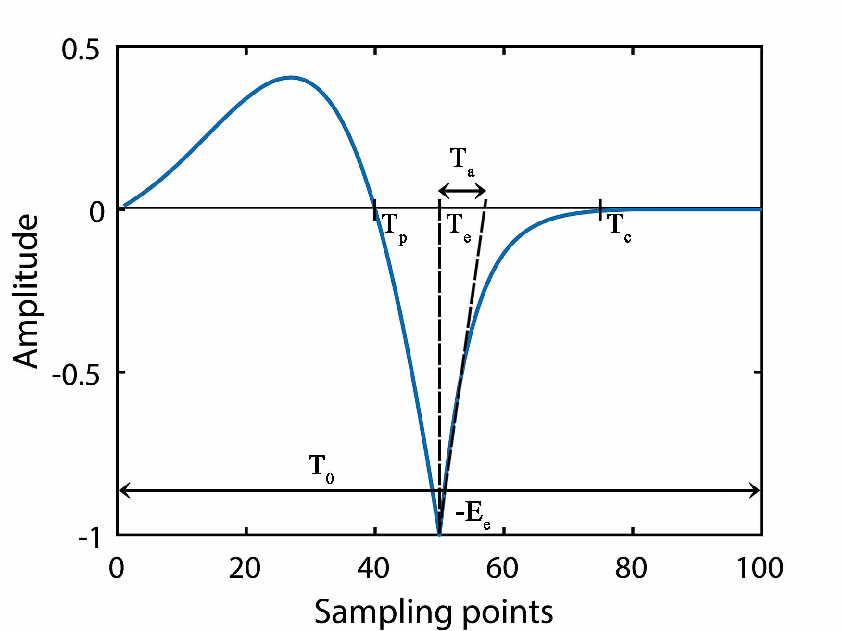}
\end{center}
\caption{Typical period of derivative of glottal source waveform represented using LF model.}
\label{fig:LF}
\end{figure}

\subsubsection{Glottal source modeled by LF}
\label{sec:GS}
The LF model proposed by Fant et al.~\cite{fant1985four} is a parametrized time-domain model for modeling the derivative of glottal source waveforms (glottal airflow). Its parameters can be approximated using inverse filtering from recorded speech. The properties of the LF model have been extensively studied. 

In the continuous time domain, a typical period of the derivative of a glottal source waveform modeled using the LF model is shown in Fig.~\ref{fig:LF}. In the LF model, six parameters ($T_{0}$, $T_{p}$, $T_{e}$, $T_{a}$, $T_{c}$ and $E_{e}$) are used to describe the shape of a derivative of a glottal source waveform~\cite{fant1985four}, where $T_{0}$ is one period of glottal flow, $T_{p}$ is the instant of the maximum glottal flow waveform, $T_{e}$ is the instant of the maximum negative differentiated glottal flow, $T_{a}$ is the duration of the return phase, $T_{c}$ is the instant at the complete glottal closure, and $E_{e}$ is the amplitude at the glottal closure instant. $T_{c}$ is often set to $T_{0}$ in a simple LF model. Five of these parameters are time dependent ($T_{0}$, $T_{p}$, $T_{e}$, $T_{a}$ and $T_{c}$) and one is amplitude related ($E_{e}$). 

Assuming the sampling frequency is $F_s$ and one period of glottal flow is $T_{0}$, the sampling period is then $T_s=1/F_s$. The LF model in the discrete-time domain for one fundamental period can be expressed as
\begin{align}
u(n)=\left\{
             \begin{array}{lr}
             E_{1}e^{\lambda nT_s}\sin(\omega n T_s), & 0\leq nT_s < T_{e} \\
             -E_2[e^{-\mu(nT_s-T_{e})}\\
             \qquad -e^{-\mu(T_{c}-T_{e})}], & T_{e} \leq nT_s < T_{c}\\
             0, &  T_{c} \leq nT_s \leq T_{0}\\
             \end{array}
\right.
\label{LF}
\end{align}

These direct synthesis parameters $\left\{E_{1}, \lambda, \mu, \omega\right\}$ can be derived with the following constraints~\cite{fant1985four}:
\begin{equation}
\begin{cases}\sum_{n=1}^{n=N_0}u(n)=0 \\\omega = \frac{\pi}{N_p}\\\mu N_{a}=1-e^{-\mu(N_c-N_e)} \\ E_1 = -\frac{E_e}{e^{\lambda N_e}\sin(\omega N_e)}\\ E_2 = \frac{E_e}{\mu N_{a}}\end{cases},
\label{constraints}
\end{equation}
where $\left\{N_{0}, N_{p}, N_{e}, N_{a}, N_{c}\right\}$ are parameters in the discrete-time domain corresponding to $\left\{T_{0}, T_{p}, T_{e}, T_{a}, T_{c}\right\}$, respectively, and can be derived as
\begin{equation}
\begin{cases}N_0 = \lfloor T_0/ T_s\rceil \\N_p=\lfloor T_p/T_s\rceil \\ N_e=\lfloor T_e/T_s\rceil\\ N_a=\lfloor T_a/T_s\rceil\\ N_c=\lfloor T_c/T_s\rceil\end{cases},
\end{equation}
where $\lfloor\cdot\rceil$ denotes the rounding function.

\subsubsection{ARX model}
Given the above assumptions, the vocal tract can be simulated using an ARX model, which combines an all-pole AR model with an additional exogenous LF excitation. In the ARX model, the speech production in the time domain can be represented as
\begin{equation}
s(n)=-\sum_{i=1}^p a_{i}s(n-i)+b_{0}u(n)+e(n),
\label{synthesis1}
\end{equation}
where $s(n)$ is the synthesized speech at time $n$, $e(n)$ is the error, $a_{i},i=1,...,p$ are the coefficients of the ARX model, $u(n)$ is the exogenous input to the filter at $n$ generated from the LF model, and $b_{0}$ is used to adjust the amplitude of the input.

As mentioned in the Introduction, the all-pole AR model in vocal tract modeling cannot estimate the position of anti-formants. In an iteration-based estimation procedure, the glottal source parameters are renewed in each loop to search for the optimal solution. As a consequence, the parameters estimated from the glottal source waveform include anti-formant information from the vocal tract, which decreases the estimation accuracy. More reasonable modeling based on the source-filter assumption is necessary to address this issue.

\subsection{Proposed ARMAX-LF model}

The ARMAX-LF model replaces the all-pole model in the ARX-LF model with a pole-zero model. In the ARMAX model, the speech production in the time domain can be represented as

\begin{equation}
s(n)=-\sum_{i=1}^p a_{i}s(n-i)+\sum_{j=0}^q b_{j}u(n-j)+e(n),
\label{synthesis}
\end{equation}

\noindent where $a_{i},i=1,...,p$ and $b_{j},j=1,...,q$ are the coefficients of the ARMAX model. In Eq.~\ref{synthesis}, $u(n)$ is deterministic and $e(n)$ is probabilistic. To estimate $a_{i}$ and $b_{j}$, we transform Eq. (\ref{synthesis}) into 
\begin{equation}
e(n)=s(n)+\sum_{i=1}^p a_{i}s(n-i)-\sum_{j=0}^q b_{j}u(n-j).
\label{en}
\end{equation}

This can also be represented in the z-domain as
\begin{equation}
E(z)=S(z)A(z)-U(z)B(z).
\label{zdomain}
\end{equation}

If we minimize the power of $e(n)$, $E(z)\rightarrow0$, then
\begin{equation}
S(z)=\frac{E(z)+U(z)B(z)}{A(z)}\rightarrow\frac{B(z)}{A(z)}U(z).
\label{zdomain}
\end{equation}

The transfer function $H(z)$ is
\begin{equation}
H(z)=\frac{B(z)}{A(z)}=\frac{\sum_{j=0}^q b_{j}z^{-j}}{\sum_{i=0}^p a_{i}z^{-i}}=\frac{\prod_{j=1}^q (1-\beta_j z^{-1})}{\prod_{i=1}^p (1-\alpha_i z^{-1})},
\label{zdomain}
\end{equation}

\noindent where $\alpha_i$ and $\beta_j$ refer to the pole and zero in the vocal tract transfer function and can be derived from $a_i$ and $b_j$. 

For convenience, $a_0$ is set to 1, which transforms Eq.~(\ref{en}) to Eq. (\ref{en1}) and into a matrix form
\begin{equation}
e(n)=\sum_{i=0}^p a_{i}s(n-i)-\sum_{j=0}^q b_{j}u(n-j)
\label{en1}
\end{equation}

\begin{equation}
\mathbf{e}=\mathbf{Sa}-\mathbf{Ub}=\begin{bmatrix}\mathbf{S}\mid -\mathbf{U} \end{bmatrix}\begin{bmatrix}\mathbf{a} \\\mathbf{b} \end{bmatrix}=\mathbf{Fh},
\end{equation}
where
\begin{equation}
\begin{split}
    &\mathbf{e}=\begin{bmatrix} e(n) \\e(n-1)\\\vdots\\e(n-N+1)\end{bmatrix},\quad \mathbf{s}_{i}=\begin{bmatrix}s(n-i) \\s(n-i-1)\\\vdots\\s(n-i-N+1)\end{bmatrix},\\
    &\mathbf{S}=[s_{0}\,s_{1}\cdots s_{p}], \quad \mathbf{u}_{j}=\begin{bmatrix}u(n-j) \\u(n-j-1)\\\vdots\\u(n-j-N+1) \end{bmatrix},\\
    &\mathbf{U}=[u_{0}\,u_{1}\cdots u_{q}], \quad \mathbf{a}=\begin{bmatrix}a_{0}\\a_{1}\\\vdots\\a_{p} \end{bmatrix},\quad \mathbf{b}=\begin{bmatrix}b_{0}\\b_{1}\\\vdots\\b_{q} \end{bmatrix},\\
    &\mathbf{F}=\begin{bmatrix}\mathbf{S}\mid -\mathbf{U} \end{bmatrix},\quad \mathbf{h}=\begin{bmatrix}\mathbf{a}\\\mathbf{b} \end{bmatrix}.
\end{split}
\end{equation}

For one period of glottal vibration, $N$ is the number of sampling points in $T_{0}$. To obtain the optimal coefficients $\mathbf{h}$ from the ARMAX model, we seek to minimize the mean-square error (MSE) $Exp(e^{\top}e)$, where $Exp(\cdot)$ denotes the mathematical expectation. Taking the gradient of $Exp(e^{\top}e)$ with respect to $a_{i}$ and $b_{j}$ and equating to $\mathbf{0_{(p+q+2)\times 1}}$, then we can use the Wiener-Hopf equation to obtain the optimal solution:

\begin{equation}
\mathbf{h}=-(\mathbf{F}^{\top}\mathbf{F})^{-1}\mathbf{F}^{\top}\mathbf{s_{0}}.
\label{finalstep}
\end{equation}

\section{Estimation method}

\begin{figure*}[t]
\centering
\includegraphics[width=0.9\linewidth]{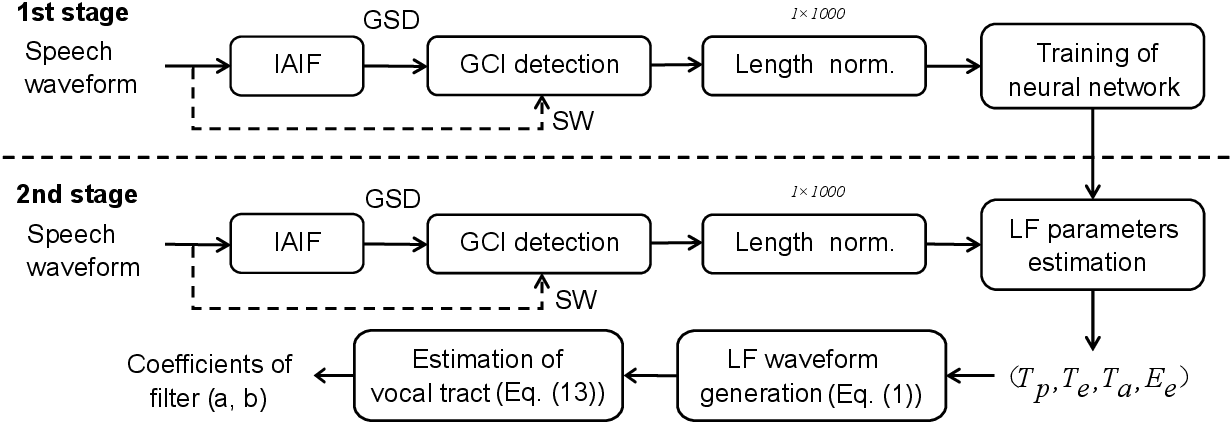}
\caption{Block diagram of proposed two-stage method for estimating parameters in glottal source derivative (GSD) and vocal tract filter. Speech waveform (SW) or GSD was used as input features to train the neural network and estimate LF parameters.}
\label{fig:estimation_procedure}
\end{figure*}

In the ARMAX-LF model, the glottal source is modeled by the derivative of the glottal source waveform using the parametrized LF model, and the vocal tract is represented using a pole-zero filter with an additional exogenous residual signal as an ARMAX model. The estimation of the ARMAX-LF model is a multi-parameter nonlinear joint-optimization problem. When we estimate the ARMAX model parameters from the recorded speech, the glottal source excitation based on the LF assumption must be known. Therefore, the estimation accuracy of glottal source parameters is crucial for estimating the vocal tract filter.

As shown in Fig.~\ref{fig:estimation_procedure}, the estimation process for the glottal source and vocal tract parameters is roughly divided into two stages. The first stage is the training of DNN. In this stage, a mapping from input features to corresponding LF parameters ($T_p$, $T_e$, $T_a$, $E_e$) is trained on the basis of the powerful nonlinear fitting ability of DNNs. The second stage is the implementation of glottal source and vocal tract parameters estimation using the trained model and Eq.~(\ref{finalstep}). In this stage, glottal source and vocal tract parameters with fewer estimation errors are estimated without any iterations as in the analysis-by-synthesis strategy.

\subsection{Training of DNN for LF model initialization}
The training procedure is shown in the first stage of the block diagram of Fig.~\ref{fig:estimation_procedure}. Inverse filtering is widely used in estimating glottal excitation. By using inverse filtering, the glottal source waveform can be obtained by canceling the effects of the vocal tract through the inverse of the transfer function of the vocal tract. It has been proven to be an efficient method for estimating the glottal source waveform. The iterative and adaptive inverse filtering (IAIF) method proposed by Alku~\cite{alku1992glottal} has become a representative method in inverse filtering~\cite{drugman2012comparative}. This method is based on an iterative process between the vocal tract and glottal source to obtain the parameters of the glottal source model. 

As introduced in Section \ref{sec:GS}, the LF model is designed to simulate the glottal source derivative (GSD) signal. It seems straightforward to use the GSD as a front-end input feature; however, estimation accuracy could be compromised due to errors in the IAIF algorithm. An alternative approach is to use the speech waveform (SW) as the input feature. Therefore, in the first stage, the SW or GSD extracted from the IAIF was used as the front-end input of the DNN model. Comparing the performance of GSD and SW helps optimize the effectiveness of the DNN-based estimation method.

Glottal closure instances (GCIs) refer to the instants of significant excitation of the vocal tract. The distance between two continuous GCIs is viewed as one period ($T_{0}$). The LF model is used to simulate the excitation signal within each period. This paper used the GCI-detection technique called speech event detection using the residual excitation and a mean-based signal (SEDREAMS)~\cite{drugman2011detection} to cut signals into different periods. Then, three contiguous periods were concatenated and padded to fix the dimension of NN input.

A feed-forward NN that has three layers of hidden units between input and output layers was used to estimate the parameters of the LF model. Estimating LF parameters is a regression problem where we are estimating discrete outputs. Therefore, we chose a linear activation function for the output (regression) layer and sigmoid activation function units for the hidden layers. The latter is defined as
\begin{equation}
y_{i}=f\left(  \sum_j W_{ij} x_j +\gamma_i \right),
\end{equation}
where $f(\alpha) = 1/(1 + \exp(-\alpha))$ is the sigmoid activation function, $W_{ij}$ and $\gamma_i$ are weights and biases, and $x_j$ and $y_i$ are the input and output, respectively. For the linear layer, the activation function is calculated as
\begin{equation}
y_{i}=\sum_jW_{ij} x_{j}+\gamma_i.
\end{equation}

The DNN is trained by back-propagating derivatives of a cost function that measures the discrepancy between the target outputs and the actual outputs. This work uses mean squared error (MSE) as the cost function. The cost function is calculated as
\begin{equation}
E=\sum_m(y_m-\hat{y}_m)^{2},
\end{equation}
where $\hat{y}_m$ is the reference value of LF model parameters for NN training. $m$ is the number of estimated parameters.

\subsection{Implementation of glottal source and vocal tract estimation}
In the second stage, the LF model parameters and vocal tract coefficients, including formants and anti-formants, are estimated with the trained model and Eq.~(\ref{finalstep}). Detailed implementation processes can be found in the second stage of Fig.~\ref{fig:estimation_procedure}. In this stage, the pre-processing used is the same as the first stage. Four glottal source parameters, $T_p$, $T_e$, $T_a$ and $E_e$, are predicted from the trained model. Then, LF waveforms are generated using Eq. \ref{LF}. Finally, coefficients of the vocal tract filter can be calculated by using Eq. \ref{finalstep}.

\section{Datasets and Experiments}

\begin{table}[t]
\centering
\caption{Architecture of DNN for glottal source modeling.}
\begin{tabular}{ccc}
\hline
Layer             & Output shape  & Params   \\ \hline
Feature extraction & {[}64, 1000{]} & -        \\ \hline
0\_FC1          & {[}64, 300{]} & 210.3k   \\
1\_BatchNorm1d   & {[}64, 300{]} & 600      \\
2\_Sigmod        & {[}64, 300{]} & -        \\
3\_FC2          & {[}64, 300{]} & 90.3k    \\
4\_BatchNorm1d       & {[}64, 300{]} & 600      \\
5\_Sigmod        & {[}64, 300{]} & -        \\
6\_FC3            & {[}64, 4{]}   & 1.505k   \\ \hline
Total                       & -             & 303.305k \\ \hline
\end{tabular}
\label{tab:architecture}
\end{table}

\begin{table*}[t]
\begin{center}
\setlength{\tabcolsep}{0.8mm}{
\caption{Formants and anti-formants frequencies (Freq.) and bandwidths (BW) of five vowels (/a/, /e/, /u/, /i/, and /o/) and two nasalized consonants (/m/ and /$\widetilde{\varepsilon}$/). Unit: Hz.}
\label{tab:sys_para}
\begin{tabular}{ccccccccccccccccccccc}
\hline
Formants &
  \multicolumn{2}{c}{/a/} &
   &
  \multicolumn{2}{c}{/i/} &
   &
  \multicolumn{2}{c}{/u/} &
   &
  \multicolumn{2}{c}{/e/} &
   &
  \multicolumn{2}{c}{/o/} &
   &
  \multicolumn{2}{c}{/m/} &
   &
  \multicolumn{2}{c}{/$\widetilde{\varepsilon}$/} \\ \cline{2-3} \cline{5-6} \cline{8-9} \cline{11-12} \cline{14-15} \cline{17-18} \cline{20-21} 
   & Freq. & BW  &  & Freq. & BW  &  & Freq. & BW  &  & Freq. & BW  &  & Freq. & BW  &  & Freq. & BW  &  & Freq. & BW  \\ \cline{1-3} \cline{5-6} \cline{8-9} \cline{11-12} \cline{14-15} \cline{17-18} \cline{20-21} 
F1 & 750   & 90  &  & 281   & 90  &  & 312   & 90  &  & 469   & 90  &  & 468   & 90  &  & 220   & 60  &  & 690   & 70  \\
F2 & 1187  & 110 &  & 2281  & 110 &  & 1219  & 110 &  & 2031  & 110 &  & 781   & 110 &  & 1050  & 100 &  & 1640  & 100 \\
F3 & 2595  & 170 &  & 3187  & 170 &  & 2469  & 170 &  & 2687  & 170 &  & 2656  & 170 &  & 2380  & 120 &  & 1940  & 110 \\
F4 & 3781  & 250 &  & 3781  & 250 &  & 3406  & 250 &  & 3375  & 250 &  & 3281  & 250 &  & 4100  & 180 &  & 2760  & 130 \\
F5 & 4200  & 300 &  & 4200  & 300 &  & 4200  & 300 &  & 4200  & 300 &  & 4200  & 300 &  & -     & -   &  & 3500  & 160 \\
F6 & -     & -   &  & -     & -   &  & -     & -   &  & -     & -   &  & -     & -   &  & -     & -   &  & 4500  & 200 \\ \hline
A1 & -     & -   &  & -     & -   &  & -     & -   &  & -     & -   &  & -     & -   &  & 1600  & 70  &  & 2260  & 250 \\
A2 & -     & -   &  & -     & -   &  & -     & -   &  & -     & -   &  & -     & -   &  & 3320  & 130 &  & -     & -   \\ \hline
\end{tabular}}
\end{center}
\end{table*}

\subsection{Datasets}

The evaluation datasets consist of synthesized speech from a linear source-filter model, synthesized speech from a non-linear physical (source-filter) model, and natural speech. This study generated both the synthesized speech from the linear source-filter model and the natural speech. Meanwhile, the synthesized speech from the physical model is sourced from OPENGLOT \cite{alku2019openglot}. Utilizing these three datasets enables a comprehensive assessment of the proposed method across matched synthetic models, unmatched synthetic models, and natural speech.

\begin{table}[t]
\begin{center}
\caption{Statistics of datasets used for iteration-based (dataset 1) and proposed (dataset 2) estimation methods.}
\label{tab:database}
\begin{tabular}{ccc}
\hline
\textbf{} & Dataset 1 & Dataset 2 \\ \hline
Vowel     & 225       & 67500     \\
Consonant & 90        & 27000     \\ \hline
Total     & 315       & 94500     \\ \hline
\end{tabular}
\end{center}
\end{table}

\begin{table*}[t]
\begin{center}
\caption{Average estimation errors ($\overline{\epsilon}$) of glottal source parameters for synthesized speech signals using DyProg with iteration and proposed methods. SW-DNN and GSD-DNN refer to proposed DNN-based estimation methods, with speech waveform and glottal source derivative as inputs.}
\label{tab:results1}
\begin{tabular}{ccccclclclcll}
\hline
\multirow{2}{*}{\textbf{}}  & \multirow{2}{*}{\textbf{Model}} & \multirow{2}{*}{\textbf{Estimation method}} & \multirow{2}{*}{\textbf{Dataset}} & \multicolumn{9}{c}{Error of glottal source parameters (\%)}                                                                        \\ \cline{5-13} 
                            &                                 &                                             &                                   &  & \multicolumn{1}{c}{Tp}    &  & \multicolumn{1}{c}{Te}    &  & \multicolumn{1}{c}{Ta}     &  & \multicolumn{2}{c}{Ee}            \\ \hline
\multirow{6}{*}{Vowels}     & ARX-LF                          & DyProg + iteration                          & 1                                 &  & \multicolumn{1}{c}{8.78}  &  & \multicolumn{1}{c}{8.24}  &  & \multicolumn{1}{c}{126.73} &  & \multicolumn{2}{c}{24.37}         \\
                            & ARMAX-LF                        & DyProg + iteration                          & 1                                 &  & \multicolumn{1}{c}{15.44} &  & \multicolumn{1}{c}{15.72} &  & \multicolumn{1}{c}{93.57}  &  & \multicolumn{2}{c}{23.10}         \\
                            & ARMAX-LF                        & GSD-DNN                                     & 1                                 &  & \textbf{1.31$\downarrow$}             &  & 1.21                      &  & \textbf{12.33$\downarrow$}             &  & \multicolumn{2}{l}{0.73}          \\
                            & ARMAX-LF                        & SW-DNN                                      & 1                                 &  & 1.44                      &  & \textbf{1.05$\downarrow$}             &  & 12.44                      &  & \multicolumn{2}{l}{\textbf{0.71$\downarrow$}} \\
                            & ARMAX-LF                        & GSD-DNN                                     & 2                                 &  & 4.13                      &  & 3.59                      &  & 23.27                      &  & \multicolumn{2}{l}{0.03}          \\
                            & ARMAX-LF                        & SW-DNN                                      & 2                                 &  & \textbf{3.96$\downarrow$}             &  & \textbf{2.83$\downarrow$}             &  & \textbf{21.28$\downarrow$}             &  & \multicolumn{2}{l}{\textbf{0.03$\downarrow$}} \\ \hline
\multirow{6}{*}{Consonants} & ARX-LF                          & DyProg + iteration                          & 1                                 &  & \multicolumn{1}{c}{13.83} &  & \multicolumn{1}{c}{13.27} &  & \multicolumn{1}{c}{167.97} &  & \multicolumn{2}{c}{26.64}         \\
                            & ARMAX-LF                        & DyProg + iteration                          & 1                                 &  & \multicolumn{1}{c}{14.02} &  & \multicolumn{1}{c}{14.54} &  & \multicolumn{1}{c}{166.73} &  & \multicolumn{2}{c}{25.48}         \\
                            & ARMAX-LF                        & GSD-DNN                                     & 1                                 &  & \textbf{1.01$\downarrow$}             &  & \textbf{0.94$\downarrow$}             &  & \textbf{8.88$\downarrow$}              &  & \multicolumn{2}{l}{\textbf{0.46$\downarrow$}} \\
                            & ARMAX-LF                        & SW-DNN                                      & 1                                 &  & 1.70                      &  & 1.38                      &  & 13.26                      &  & \multicolumn{2}{l}{0.91}          \\
                            & ARMAX-LF                        & GSD-DNN                                     & 2                                 &  & 5.11                      &  & 4.62                      &  & 23.31                      &  & \multicolumn{2}{l}{0.03}          \\
                            & ARMAX-LF                        & SW-DNN                                      & 2                                 &  & \textbf{4.99$\downarrow$}             &  & \textbf{3.41$\downarrow$}             &  & \textbf{21.92$\downarrow$}             &  & \multicolumn{2}{l}{\textbf{0.03$\downarrow$}} \\ \hline
\end{tabular}
\end{center}
\end{table*}

\subsubsection{Synthesized speech using source-filter model}

Synthesized vowels (/a/, /e/, /i/, /o/, and /u/) were obtained using Kawahara's method~\cite{kawahara2016sparkng}. To obtain synthesized consonants, we replaced the AR model in Kawahara's method with an ARMA filter in the vocal tract modeling to synthesize nasalized consonants such as /m/ and /$\widetilde{\varepsilon}$/. The detailed positions of formants and anti-formants for each synthetic speech are listed in Table \ref{tab:sys_para}. For each syllable, the LF model parameters were varied: $T_e$: 0.3 to 0.9; $T_p/T_e$: 0.65 to 0.8; and $T_a$: 0.03 to 0.08, within the range suggested in~\cite{drugman2012comparative}. $T_c$ and $E_e$ were fixed to 1 following \cite{li2020simultaneous}. To synthesize speech that covers more realistic pronunciation circumstances, a large number of vowels and consonants were synthesized with 45 $F_{0}$s varied from 80 to 300 Hz. Three hundred combinations of LF model parameters were selected randomly for each syllable and $F_{0}$. Finally, the synthesized vowels and synthesized consonants had 67500 ($300 \times 45 \times 5$) and 27000 ($300 \times 45 \times 2$) different conditions, respectively. Because estimation using the iteration-based method is very time-consuming, we separate all synthesized speech into a small dataset (dataset 1) and a big dataset (dataset 2) with the same synthesis method. The LF model parameters of dataset 1 are fixed. The synthesized vowels and synthesized consonants of dataset 1 have 225 ($1 \times 45 \times 5$) and 90 ($1 \times 45 \times 2$) different conditions, respectively. The statistics of these two datasets are listed in Table~\ref{tab:database}. The sampling frequency of synthesized speech is 16,000 Hz, and the orders of the poles and the zeros were set to be twice the number of formants and anti-formants, respectively. The length of the synthesized speech was set to 1 s.

\subsubsection{Synthesized speech using physical model}
The OPENGLO datasets proposed by Alku {\it{et al.}} \cite{alku2019openglot} are used to objectively evaluate the effectiveness of the proposed ARMAX-LF model and DNN-based two-stage estimation procedure under unmatched model assumptions. The OPENGLOT dataset, which consists of four repositories according to the different data generation modes, aims to evaluate the performance of glottal inverse filtering. The repositories II and III are selected in this paper. Repository II consists of 93 audios of /a/, /e/, /i/, and /u/ with $F_0$ ranging from 100 to 360 Hz, and each audio contains synthetic glottal flow, glottal area, and speech pressure signal from physical modeling of speech production. Repository III consists of 287 audios of /a/, /e/, /i/, and /u/ with $F_0$ ranging from 100 to 500 Hz, and each audio contains synthetic glottal flow and speech pressure signal from a physical system with an acoustic source and 3D printed vocal track. Interested readers can get more information about data synthesis from \cite{alku2019openglot}.  

\begin{figure}[tb]
\begin{center}
\includegraphics[width=1\linewidth]{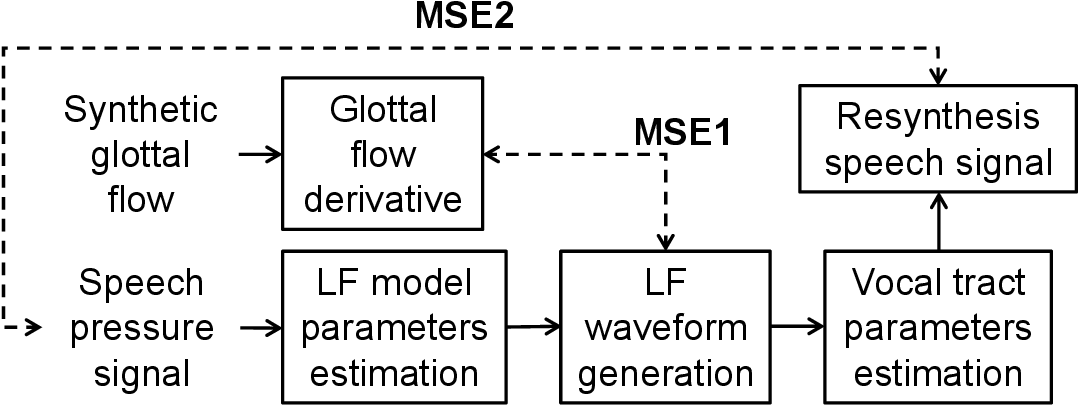}
\end{center}
\caption{The calculation method of mean square errors using the OPENGLOT datasets.}
\label{fig:mse}
\end{figure}

The repositories II and III contain the ground-truth values for synthetic glottal flow and speech pressure signals. As shown in Fig. \ref{fig:mse}, this study computes the Mean Squared Error 1 (MSE1) between the predicted and synthetic glottal flow, as well as the Mean Squared Error 2 (MSE2) between the re-synthesized and original speech pressure signals using the estimated glottal source and vocal tract parameters. MSE1 assesses the performance of the glottal source estimation methods, while MSE2 evaluates the performance of both the glottal source and vocal tract estimation methods.

\begin{figure}[!htb]
\begin{center}
\includegraphics[width=1\linewidth]{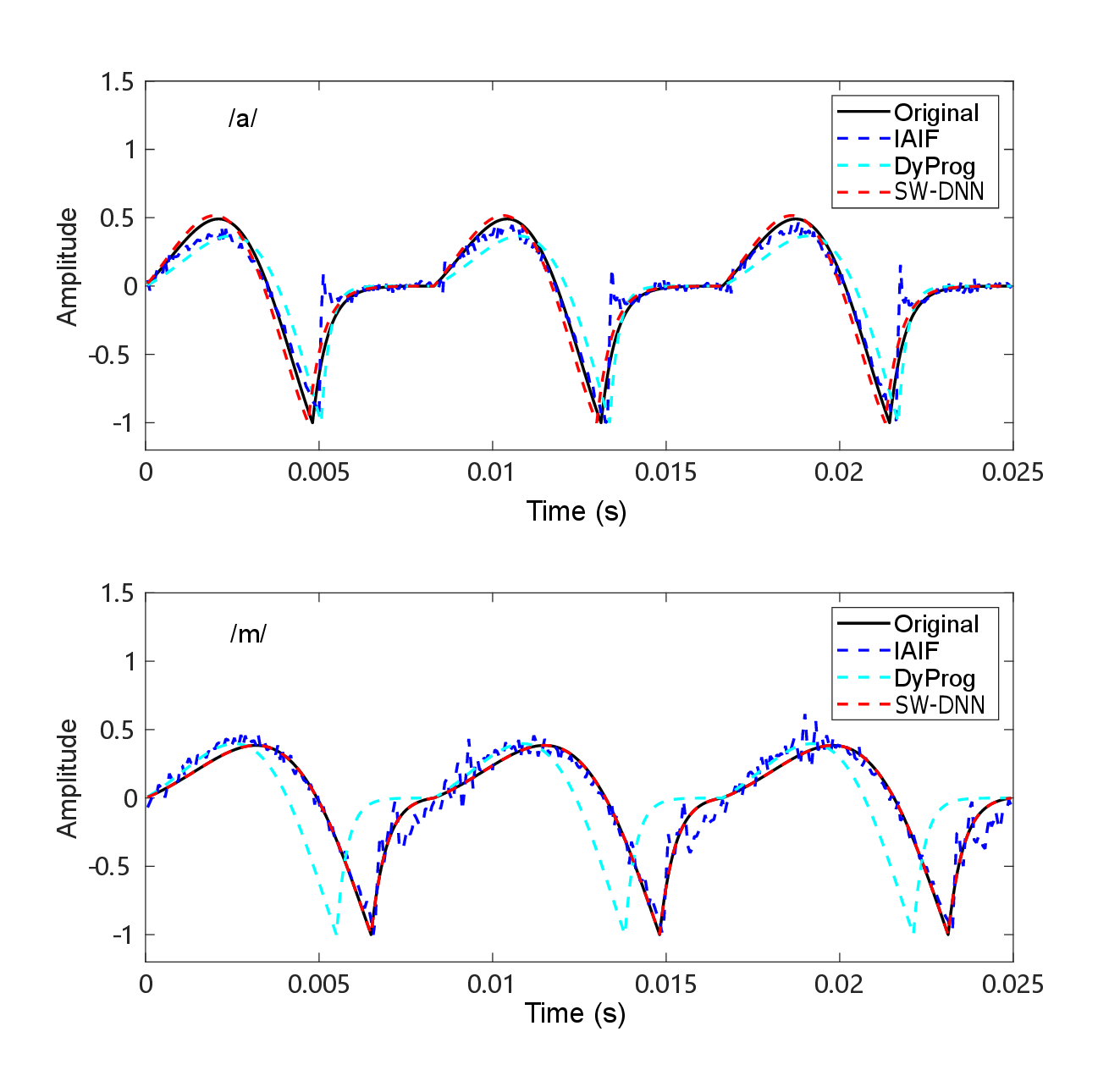}
\end{center}
\caption{Comparison of two examples in the estimation of glottal source derivative (GSD), including vowel $/a/$ (top) and consonant $/m/$ (bottom). Each subfigure includes an original GSD (black line), an estimated GSD using IAIF (blue line), two resynthesized GSDs using parameters estimated from DyProg (cyan line) and the proposed method (red line).}
\label{fig:examples1}
\end{figure}

\subsubsection{Natural speech}

Obtaining the real parameters from natural speech is impossible. Therefore, it is difficult to evaluate the estimation accuracy by creating a large database with natural speech. In this study, we recorded one syllable (/m/) pronounced by a male speaker ten times. The pronunciation conditions, such as the volume and pitch, were kept largely the same each time. We aim to show the variation range (stability) of the estimated formant and anti-formant frequencies by using the proposed method. The speech was recorded in a soundproof room at the JAIST AIS laboratory using the Audacity software. All speech was produced by the same speaker at a 44,100-Hz sampling rate in a 16-bit, mono-quality format. Downsampling was performed to decrease the sampling rate to 16,000 Hz. The duration of each utterance was restricted to 2 s. To maintain the stability of pronunciation, we selected a 1-s utterance from the middle part of each utterance.

In addition, we recorded continuous speech (/aiueo/) produced by a male speaker. We aim to show the estimation accuracy in continuous speech by comparing the similarity of the waveform and spectrogram between the recorded speech and resynthesized speech.


\subsection{Experimental setup}

The number of layers and hidden units was set to 3 and 300, respectively, following \cite{raitio2014voice}. Detailed compile and trainable parameters are listed in Table~\ref{tab:architecture}. The dimension of the input feature is 1000. The output dimension of the last fully-connected layer is 4, which corresponds to glottal source parameters ($T_p$, $T_e$, $T_a$, $E_e$). We trained the model for 30 epochs using the Adam optimizer \cite{kingma2014adam} with a learning rate of 0.01 and a batch size of 64.

The average error rate (AER) $\overline{\epsilon}$ calculated by Eq. (\ref{er}) is used to evaluate the distance between the reference and estimation values.
\begin{equation}
\overline{\epsilon}=\frac{1}{K} \sum_{i=1}^K \frac{|\widehat{\xi_i}-\xi_i|}{\xi_i},
\label{er}
\end{equation}

\noindent where $\widehat{\xi_i}$ and $\xi_i$ refer to the estimated and true value of $i$th estimated point, respectively. $K$ is the total number of estimated points.

To balance accuracy and computational complexity during the estimation process, the orders of the vocal tract filter were uniformly set to 14 for poles and 8 for zeros.

\begin{table*}[t]
\begin{center}
\caption{Average estimation error rates (AER) of vocal tract parameters for synthesized speech signals using DyProg with iteration and proposed methods. SW-DNN and GSD-DNN refer to proposed DNN-based estimation methods with speech waveform and glottal source derivative as inputs, respectively.}
\label{tab:results2}
\begin{tabular}{cccccccccc}
\hline
\multirow{2}{*}{\textbf{}}  & \multirow{2}{*}{\textbf{Model}} & \multirow{2}{*}{\textbf{Estimation method}} & \multirow{2}{*}{\textbf{Dataset}} & \multicolumn{6}{c}{Error of vocal tract parameters (\%)}                         \\ \cline{5-10} 
                            &                                 &                                             &                                   & F1            & F2            & F3            &  & A1            & A2            \\ \hline
\multirow{6}{*}{Vowels}     & ARX-LF                          & DyProg + iteration                          & 1                                 & 3.34          & 1.81          & 2.21          &  & /             & /             \\
                            & ARMAX-LF                        & DyProg + iteration                          & 1                                 & 4.01          & 0.83          & 0.55          &  & /             & /             \\
                            & ARMAX-LF                        & GSD-DNN                                     & 1                                 & \textbf{1.01$\downarrow$} & 0.41          & \textbf{0.37$\downarrow$} &  & /             & /             \\
                            & ARMAX-LF                        & SW-DNN                                      & 1                                 & 1.04          & \textbf{0.39$\downarrow$} & 0.39          &  & /             & /             \\
                            & ARMAX-LF                        & GSD-DNN                                     & 2                                 & \textbf{2.14$\downarrow$} & 0.76          & 0.59          &  & /             & /             \\
                            & ARMAX-LF                        & SW-DNN                                      & 2                                 & 2.25          & \textbf{0.71$\downarrow$} & \textbf{0.55$\downarrow$} &  & /             & /             \\ \hline
\multirow{6}{*}{Consonants} & ARX-LF                          & DyProg + iteration                          & 1                                 & 6.48          & 3.27          & 3.96          &  & /             & /             \\
                            & ARMAX-LF                        & DyProg + iteration                          & 1                                 & 6.26          & 1.36          & 1.29          &  & 6.90          & 3.41          \\
                            & ARMAX-LF                        & GSD-DNN                                     & 1                                 & \textbf{1.02$\downarrow$} & \textbf{0.79$\downarrow$} & 0.66          &  & \textbf{2.60$\downarrow$} & \textbf{4.04$\downarrow$} \\
                            & ARMAX-LF                        & SW-DNN                                      & 1                                 & 1.24          & 0.96          & \textbf{0.62$\downarrow$} &  & 3.29          & 4.68          \\
                            & ARMAX-LF                        & GSD-DNN                                     & 2                                 & \textbf{3.32$\downarrow$} & 1.98          & 0.99          &  & \textbf{5.86$\downarrow$} & 5.06          \\
                            & ARMAX-LF                        & SW-DNN                                      & 2                                 & 3.48          & \textbf{1.97$\downarrow$} & \textbf{0.98$\downarrow$} &  & 5.87          & \textbf{4.50$\downarrow$} \\ \hline
\end{tabular}
\end{center}
\end{table*}


\begin{figure*}[htb]
\begin{center}
\includegraphics[width=1\linewidth]{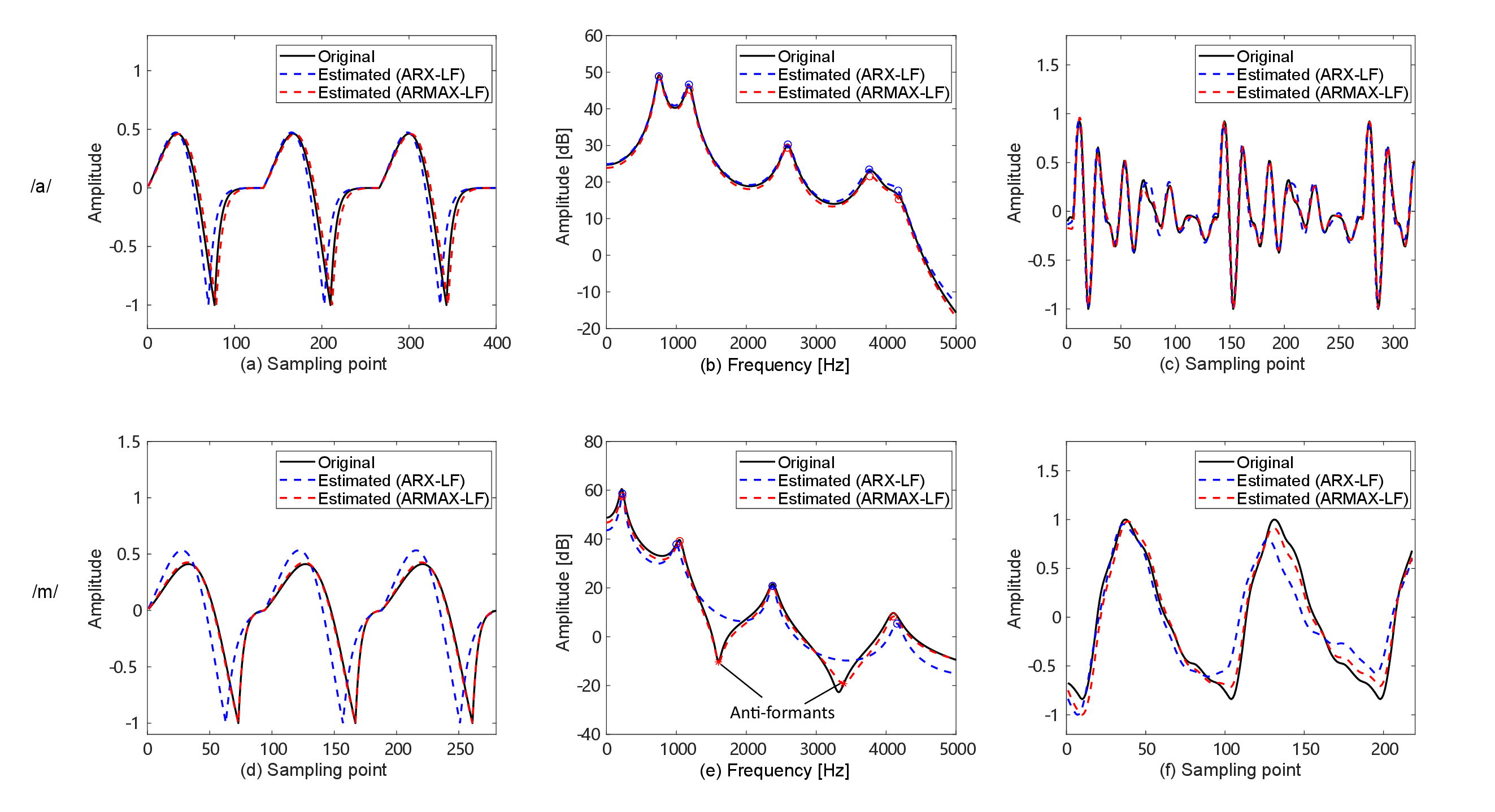}
\end{center}
\caption{Comparison of original and estimated glottal source waveforms in the time domain ((a) and (d)), the frequency response of vocal tract transfer function ((b) and (e)), and speech waveforms using the ARX-LF model with DyProg + iteration (blue line) and the ARMAX-LF model with SW-DNN (red line) estimation methods ((c) and (f)). Subfigures in the first and second lines are results estimated from synthetic vowel $/a/$ and synthetic nasalized consonant $/m/$, respectively.}
\label{fig:examples2}
\end{figure*}

\section{Results and discussion}
Dynamic programming (DyProg) is a useful technique for solving optimization problems and can be applied to estimate the glottal source parameters \cite{kane2013automating,kane2012exploiting}. This method is often used to obtain the initial values of the glottal source model, and parameters with fewer estimation errors are then estimated using an iteration-based method under the analysis-by-synthesis strategy. This paper selects the DyProg and iteration-based estimation method as baselines as in \cite{li2020simultaneous}.
\subsection{Results on glottal source estimation with synthesized speech from source-filter model}
The GSDs estimated from synthesized vowels (/a/) and nasalized speech (/m/) shown in Fig.~\ref{fig:examples1} demonstrate the effectiveness of the proposed glottal source estimation method. The original GSD (black line) is generated using Eq.~(\ref{LF}) with the LF parameters used for the original synthesis. As can be seen, the resynthesized GSD based on parameters estimated from the DyProg method yields acceptable results for vowel /a/, but the estimation accuracy decreased significantly in consonants. In contrast, the resynthesized GSD based on parameters estimated from our proposed method (SW-DNN, red line) can obtain more precise results for both vowel /a/ and consonant /m/.

We compare the AERs of DyProg + iteration and our proposed estimation methods for datasets 1 and 2. As seen in Table~\ref{tab:results1}, the ARMAX-LF model with the DyProg + iteration method increased the AER of $T_p$ and $T_e$ for both vowels and consonants. The AER in the estimation of $T_a$ is much higher than other parameters. This is because the value of $T_a$ is quite small (0.03-0.08), a smaller estimate value in a larger relative error in the estimation. The DNN-based methods can decrease errors in all LF parameters, and the improvement is significant; this indicates that the DNN is more effective than mathematical methods in estimating glottal source parameters due to its powerful nonlinear fitting ability. 

Consonants contain more errors in glottal source parameters than that of vowels, which means parameter estimation in consonants is much more difficult. It is worth noting that the error in $E_e$ is reduced to almost zero. This is because we fixed the value of $E_e$ to 1 in the synthesis stage. Fixed estimation values make it much easier for NN to construct the mapping from input features to actual values. In addition, the proposed DNN-based method with the SW feature as input performs better than with GSD input in dataset 2.

\begin{figure*}[t]
\begin{center}
\includegraphics[width=1.05\linewidth]{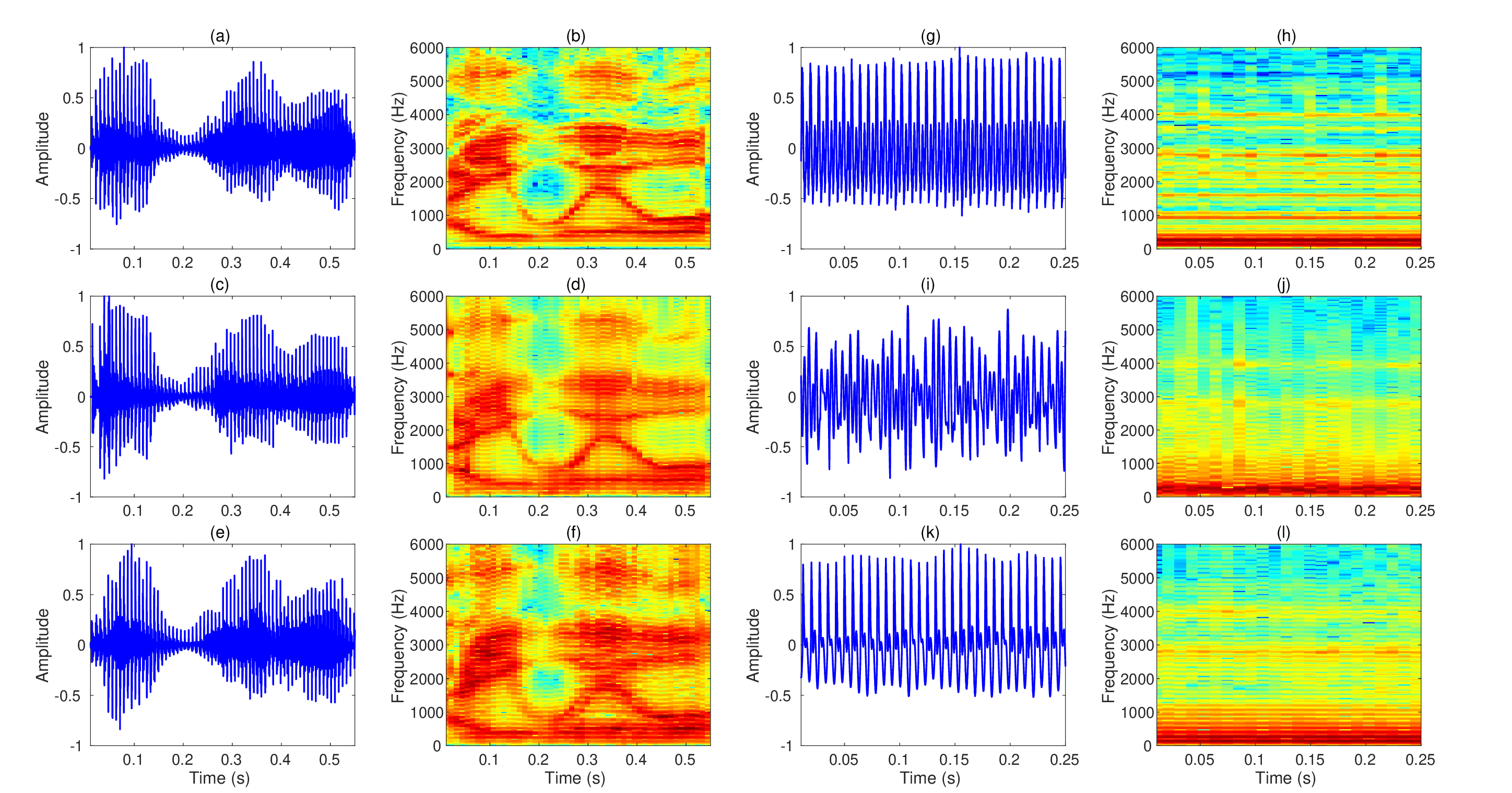}
\end{center}
\caption{Original speech waveform ((a), (g)) and its spectrogram ((b), (h)); resynthesized speech waveform ((c), (i)) and its spectrogram ((d), (j)) using the ARX-LF model with DyProg + iteration estimation method; and resynthesized speech waveform ((e), (k)) and its spectrogram ((f), (l)) using the ARMAX-LF model with SW-DNN estimation method. Subfigures (a), (b), (c), (d), (e), and (f) are the results of continuous speech (/aiueo/). Subfigures (g), (h), (i), (j), (k), and (l) are the results of natural speech (/m/).}
\label{fig:examples5}
\end{figure*}

\subsection{Results on vocal tract estimation with synthesized speech from source-filter model}
The overall results for the estimation of vocal tract parameters are listed in Table \ref{tab:results2}. F1, F2, and F3 refer to the first, second, and third formants, and A1 and A2 refer to the first and second anti-formants, respectively. The improved results are highlighted in the table. The ARMAX-LF model can estimate the position of anti-formants from consonants. However, since anti-formants are not used in vowel synthesis, the estimated anti-formants from the ARMAX-LF model are denoted as '/'.

The overall estimation errors of consonants are higher than that in vowels. Estimation errors decrease when the frequency of formants increases because a higher frequency corresponds to a bigger value of the denominator in Eq. (\ref{er}). The ARMAX-LF model outperforms the ARX-LF model in the second and third formants (F2 and F3) of both vowels and consonants when the DyProg and iteration method is used. However, the error increased in the first formant (F1) of vowels (from 3.34 \% to 4.01 \%). Estimations errors significantly decrease in all formants when the DNN-based method is used. These results indicate that accurate estimation for the glottal source parameters using the DNN-based method can significantly decrease errors in the estimation of vocal tract parameters without any iteration. The estimation errors in dataset 2 are greater than those in dataset 1. This disparity can be attributed to the presence of more complex conditions during the synthesis of the glottal source signal in dataset 2. 

Figure \ref{fig:examples2} shows the comparison of two estimation examples by using the ARX-LF model with DyProg + iteration (blue line) and the ARMAX-LF model with SW-DNN (red line) estimation method. As shown in Fig. \ref{fig:examples2}, the estimated glottal source derivatives ((a) and (d)), the frequency response of the vocal tract transfer function ((b) and (e)), and speech waveforms ((c) and (f)) using the proposed model and estimation method are more similar to the original ones. By using the proposed ARMAX-LF model, the positions of anti-formants can be clearly estimated, as shown in subfigure (e). These results show that the ARMAX-LF is more accurate than the ARX-LF, which only works well in vowels and cannot provide the positions of anti-formants. In contrast, the ARMAX-LF can be applied to consonants as well as vowels and provides accurate positions of anti-formants by assuming pole-zero characteristics in the vocal tract transfer function.

\begin{table}[tb]
\begin{center}
\caption{Comparison of estimation error (MSE) on repositories II and III from the OPENGLOT datasets. MSE1 refers to the estimation error between the glottal flow derivative and the estimated glottal flow derivative. MSE2 refers to the estimation error between the speech pressure signal and the re-synthetic speech signal.}
\label{tab:results_physical}
\begin{tabular}{cccc}
\hline
\multirow{2}{*}{\textbf{Dataset}} & \multirow{2}{*}{\textbf{\begin{tabular}[c]{@{}c@{}}Estimation\\ method\end{tabular}}} & \multirow{2}{*}{\textbf{MSE1}} & \multirow{2}{*}{\textbf{MSE2}} \\
                                  &                                                                                       &                                &                                \\ \hline
\multirow{2}{*}{Repository II}    & DyProg                                                                                & 5.86                           & 4.07                           \\ \cline{2-4} 
                                  & SW-DNN                                                                                & 8.80                           & 5.58                           \\ \hline
\multirow{2}{*}{Repository III}   & DyProg                                                                                & 10.29                          & 5.41                           \\ \cline{2-4} 
                                  & SW-DNN                                                                                & \textbf{8.52$\downarrow$}                           & \textbf{3.27$\downarrow$}                           \\ \hline
\end{tabular}
\end{center}
\end{table}

\subsection{Results on synthetic speech from physical model}

Table \ref{tab:results_physical} presents the evaluation results for synthetic speech generated using physical models. In the estimation process, both the glottal source and vocal tract are modeled using the proposed ARMAX-LF model. The performance comparison includes two estimation methods: DyProg and SW-DNN.

Compared to the DyProg method, the proposed DNN-based estimation method demonstrates improved performance, reducing MSE1 and MSE2 in Repository III. This indicates that the ARMAX-LF model can effectively handle speech generated from a mismatched physical model. In Repository II, the SW-DNN method performs worse than the DyProg method. This difference can be attributed to the varying durations of speech in Repository III (0.2s) compared to Repository II (1s). The DyProg method, which is based on an exhaustive search followed by dynamic programming, exhibits advantages in estimating longer duration speech at the expense of longer computation time. 


\subsection{Results on natural speech}

\begin{figure}[!htbp]
\begin{center}
\includegraphics[width=\linewidth]{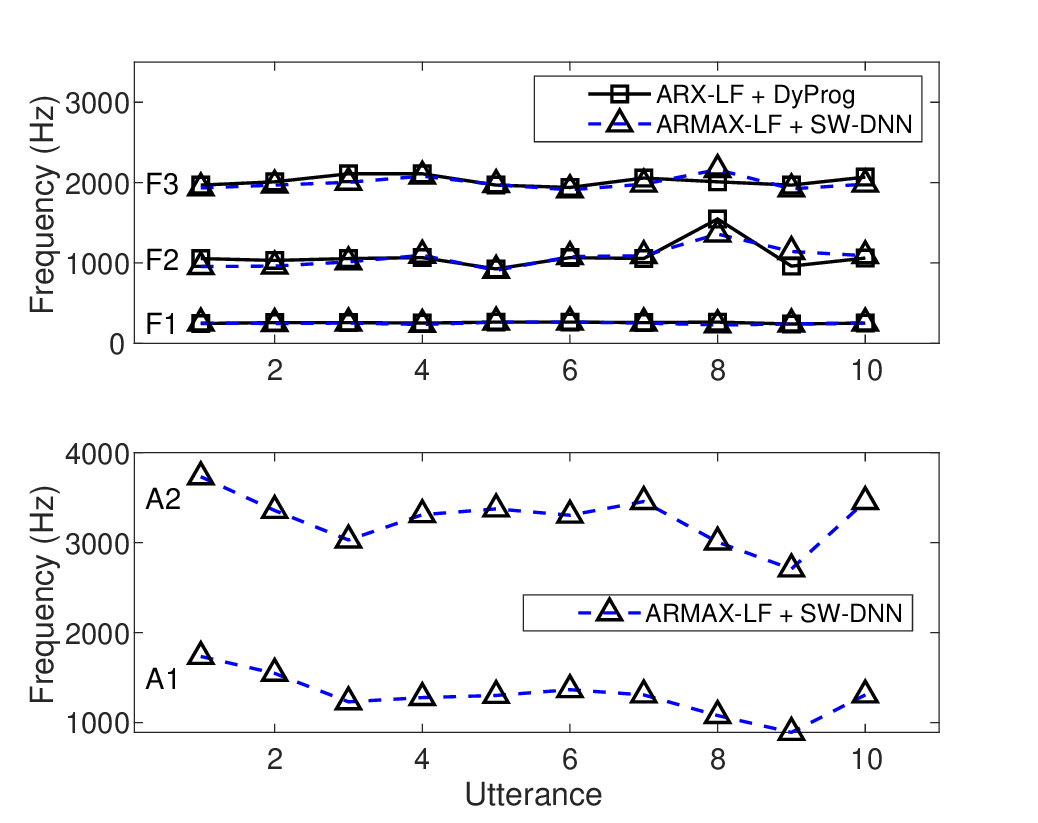}
\end{center}
\caption{Estimated frequencies of first formant (F1), second formant (F2), third formant (F3), first anti-formant (A1), and second anti-formant (A2)) using the ARX-LF model with the DyProg estimation methods and ARMAX-LF model with SW-DNN estimation method.}
\label{fig:examples4}
\end{figure}

The waveform and the spectrogram of the original speech signals (/aiueo/, /m/) and the resynthesized speech signals using the proposed method and the ARX-LF model with DyProg + iteration estimation method are plotted in Fig. \ref{fig:examples5}. For continuous natural speech /aiueo/, the resynthesized speech was more similar to the original speech in both the time and frequency domains for the two methods. However, for nasalized natural speech /m/, the accuracy of the ARX-LF model with the DyProg + iteration estimation method decreased significantly ((i) and (j)). The proposed method can obtain smaller estimation errors in the time domain (k) and more stable spectral envelopes in the frequency domain (i), which is more obvious in the frequencies of 3,000 Hz and 4,000 Hz.


The estimated positions of formants and anti-formants from recorded speech /m/ are illustrated in Fig. \ref{fig:examples4}. We can easily find that the frequency variation of the first three formants is minimal, especially in the first formant (F1). The estimated positions of the first three formants are quite near when using the ARX-LF + Dyprog and ARMAX-LF + SW-DNN methods. The frequency variation of the anti-formants is larger than that of the formants, and the variation becomes more significant as the frequency increases.

\begin{table*}[htb]
\begin{center}
\caption{Comparison of estimation time between ARX-LF model with DyProg + iteration estimation method and proposed ARMAX-LF model with SW-DNN estimation method. Unit: second (s).}
\label{tab:time}
\begin{tabular}{ccccc}
\hline
Model    & Estimation methods & Estimation duration (s) & Periods & Estimation time (s) \\ \hline
ARX-LF   & DyProg + iteration & 0.58                    & 71      & 146.77              \\
ARMAX-LF & SW-DNN             & 0.58                    & 71      & \textbf{2.33$\downarrow$}       \\ \hline
\end{tabular}
\end{center}
\end{table*}

\subsection{Results on estimation time}
A short estimation time is crucial for applying physiological parameters derived from speech production. The estimation duration, estimated periods, and estimation time of natural speech are listed in Table~\ref{tab:time}. The estimation time includes the entire process of estimating the glottal source and vocal tract parameters. The proposed ARMAX-LF model with the SW-DNN estimation method can significantly decrease the estimation time. 

\subsection{Discussion}
The proposed ARMAX-LF model and two-stage DNN-based estimation procedure outperformed the other methods in the estimation accuracy of the glottal source and vocal tract parameters for several reasons. First, the proposed ARMAX-LF model is more generalized than the ARX-LF model for modeling speech production due to the pole-zero assumption in the vocal tract transfer function. The pole-zero assumption can provide positions in both formants and anti-formants, so it can be applied to a wider variety of speech sounds, including vowels and nasalized consonants. The second reason is that parameters from the ARMAX and LF models are estimated using a two-stage estimation method, which removes the effect of the glottal source when estimating the vocal tract. Lastly, the powerful nonlinear fitting ability of the DNN is utilized instead of mathematical methods to provide accurate values of LF parameters. With LF parameters, our proposed estimation procedure can be used to estimate vocal tract parameters without any iterations and updates, significantly reducing the computation and estimation time.

Estimating the glottal source and vocal tract parameters (including formants and anti-formants) from raw speech can have several important applications in speech analysis. These parameters can provide a large amount of speaker individual information due to the physiological and morphological differences of speech organs among speakers. Based on this, one possible application is to use these parameters as features in a speaker recognition system \cite{becker2008forensic,enzinger2011logarithmic}. For example, Ewald et al. \cite{enzinger2011logarithmic} proposed using the formants and anti-formants of nasal consonants for speaker verification and achieved comparable results to the mel-frequency cepstral coefficient (MFCC) feature while offering more interpretations in terms of speech production. Speaker anonymization is another possible application by revising speaker individuality concealed in the glottal source and vocal tract parameters. For example, shifting the frequencies of formants using the McAdams coefficient has been successfully applied as a baseline system in the VoicePrivacy Challenge 2022 \cite{tomashenko2022voiceprivacy} to conceal a speaker's identity. However, the formants used in \cite{tomashenko2022voiceprivacy} are derived from the LP coefficients of the speech signal, not a physiological vocal tract. Shifting the frequencies of formants and anti-formants estimated as in this paper can potentially be used to conceal the speaker's identity while decreasing the distortion in resynthesized speech quality. In addition, accurate glottal source and vocal tract parameters can also be used in emotional speech recognition \cite{li2018contributions} and provide important information for diagnosing and treating speech disorders and diseases, such as dysphonia or vocal fold paralysis \cite{alku2011glottal,drugman2014glottal}.

However, there are also several drawbacks associated with this model, the main one being its limited flexibility. The LF model is a parametric model that represents the glottal source as a time-domain waveform generated by a set of parameters. It assumes that the glottal source derivative switches instantaneously between two smooth functions at a time $T_e$, which severely violates the Nyquist theorem and causes aliasing. In addition, a fixed-order filter may not be optimal for all types of speech signals, such as unvoiced speech.
Finally, the LF model is sensitive to noise in the speech signal, which can lead to inaccurate estimation of the glottal source parameters. This is because the model relies on the precise identification of GCI, which may be difficult in the presence of noise. The discussion of the robustness of the proposed method will be our future work.



\section{Conclusion}
We proposed the ARMAX-LF model to approximatively represent the glottal source derivative and vocal tract filter and extend the ARX-LF model to a wider variety of speech sounds. The ARMAX model represents the vocal tract as a pole-zero filter with an additional exogenous LF excitation to provide the locations of anti-formants. The LF model represents the glottal source waveform as a parametrized time-domain model. In addition, we proposed a two-stage estimation procedure based on the DNN to solve this multi-parameter nonlinear joint-optimization problem. In the first stage, a mapping from input features to corresponding LF parameters ($T_p$, $T_e$, $T_a$, $E_e$) is trained using the powerful nonlinear fitting ability of DNNs. In the second stage, the glottal source and vocal tract parameters are estimated accurately without any iterations using the trained model with the ARMAX-LF model. 

We tested the ARMAX-LF model with the two-stage DNN-based estimation method and the ARX-LF model with the DyprogLF + iteration estimation method on synthesized vowels and nasalized consonants using the linear source-filter and physical models respectively. The results for the synthesized speech show that the ARMAX-LF model can estimate the positions of anti-formants, which is impossible for the ARX-LF model. By combining the ARMAX-LF model with the two-stage DNN-based estimation method, the AERs significantly decreased in the parameters from the glottal source and vocal tract models. This effectiveness can also be verified when testing with data synthesized from physical models. The evaluation of continuous speech also shows a high similarity between natural and resynthesized speech in both time and frequency domains. In addition, the estimation time of natural speech is significantly reduced, which makes it possible to apply physiological parameters of speech production. In general, the proposed model and estimation methods demonstrate strong performance across speech generated by matched models, unmatched models, and the human vocal system. Future work will focus on applying the estimated parameters to speaker-related applications, such as speaker recognition and speaker anonymization.

\section{Acknowledgment}
This work was supported by SCOPE Program of Ministry of Internal Affairs and Communications (Grant Number: 201605002), a Grant-in-Aid for Scientific Research (Grant number: 20H04207), a Grant-in-Aid for Scientific Research (Grant number: 21H03463), and the Fund for the Promotion of Joint International Research (Fostering Joint International Research (B))(20KK0233).

\bibliographystyle{IEEEtran}
\bibliography{mybib}

\begin{thebibliography}{10}
\providecommand{\url}[1]{#1}
\csname url@samestyle\endcsname
\providecommand{\newblock}{\relax}
\providecommand{\bibinfo}[2]{#2}
\providecommand{\BIBentrySTDinterwordspacing}{\spaceskip=0pt\relax}
\providecommand{\BIBentryALTinterwordstretchfactor}{4}
\providecommand{\BIBentryALTinterwordspacing}{\spaceskip=\fontdimen2\font plus
\BIBentryALTinterwordstretchfactor\fontdimen3\font minus \fontdimen4\font\relax}
\providecommand{\BIBforeignlanguage}[2]{{%
\expandafter\ifx\csname l@#1\endcsname\relax
\typeout{** WARNING: IEEEtran.bst: No hyphenation pattern has been}%
\typeout{** loaded for the language `#1'. Using the pattern for}%
\typeout{** the default language instead.}%
\else
\language=\csname l@#1\endcsname
\fi
#2}}
\providecommand{\BIBdecl}{\relax}
\BIBdecl

\bibitem{enzinger2011speaker}
E.~Enzinger and P.~Balazs, ``Speaker verification using pole/zero estimates of nasals,'' \emph{Analele Universitatii “Eftimie Murgu}, vol.~18, pp. 33--44, 2011.

\bibitem{enzinger2011logarithmic}
E.~Enzinger, P.~Balazs, D.~Marelli, and T.~Becker, ``A logarithmic based pole-zero vocal tract model estimation for speaker verification,'' in \emph{2011 IEEE International Conference on Acoustics, Speech and Signal Processing (ICASSP)}.\hskip 1em plus 0.5em minus 0.4em\relax IEEE, 2011, pp. 4820--4823.

\bibitem{li2017commonalities}
Y.~Li, K.-I. Sakakibara, D.~Morikawa, and M.~Akagi, ``Commonalities of glottal sources and vocal tract shapes among speakers in emotional speech,'' in \emph{International Seminar on Speech Production}.\hskip 1em plus 0.5em minus 0.4em\relax Springer, 2017, pp. 24--34.

\bibitem{li2018contributions}
Y.~Li, J.~Li, and M.~Akagi, ``Contributions of the glottal source and vocal tract cues to emotional vowel perception in the valence-arousal space,'' \emph{The Journal of the Acoustical Society of America}, vol. 144, no.~2, pp. 908--916, 2018.

\bibitem{perrotin2020glottal}
O.~Perrotin and I.~V. McLoughlin, ``Glottal flow synthesis for whisper-to-speech conversion,'' \emph{IEEE/ACM Transactions on Audio, Speech, and Language Processing}, vol.~28, pp. 889--900, 2020.

\bibitem{juvela2019glotnet}
L.~Juvela, B.~Bollepalli, V.~Tsiaras, and P.~Alku, ``Glotnet—a raw waveform model for the glottal excitation in statistical parametric speech synthesis,'' \emph{IEEE/ACM Transactions on Audio, Speech, and Language Processing}, vol.~27, no.~6, pp. 1019--1030, 2019.

\bibitem{keller1995fundamentals}
E.~Keller, \emph{Fundamentals of speech synthesis and speech recognition: basic concepts, state-of-the-art and future challenges}.\hskip 1em plus 0.5em minus 0.4em\relax John Wiley and Sons Ltd., 1995.

\bibitem{wang2019neural}
X.~Wang, S.~Takaki, and J.~Yamagishi, ``Neural source-filter waveform models for statistical parametric speech synthesis,'' \emph{IEEE/ACM Transactions on Audio, Speech, and Language Processing}, vol.~28, pp. 402--415, 2019.

\bibitem{rao2018psfm}
A.~Rao and P.~K. Ghosh, ``Psfm—a probabilistic source filter model for noise robust glottal closure instant detection,'' \emph{IEEE/ACM Transactions on Audio, Speech, and Language Processing}, vol.~26, no.~9, pp. 1645--1657, 2018.

\bibitem{titze2016sensitivity}
I.~R. Titze and A.~Palaparthi, ``Sensitivity of source--filter interaction to specific vocal tract shapes,'' \emph{IEEE/ACM Transactions on Audio, Speech, and Language Processing}, vol.~24, no.~12, pp. 2507--2515, 2016.

\bibitem{rao2018glottal}
A.~Rao and P.~K. Ghosh, ``Glottal inverse filtering using probabilistic weighted linear prediction,'' \emph{IEEE/ACM Transactions on Audio, Speech, and Language Processing}, vol.~27, no.~1, pp. 114--124, 2018.

\bibitem{atal1971speech}
B.~S. Atal and S.~L. Hanauer, ``Speech analysis and synthesis by linear prediction of the speech wave,'' \emph{The journal of the acoustical society of America}, vol.~50, no.~2B, pp. 637--655, 1971.

\bibitem{makhoul1975linear}
J.~Makhoul, ``Linear prediction: A tutorial review,'' \emph{Proceedings of the IEEE}, vol.~63, no.~4, pp. 561--580, 1975.

\bibitem{makhoul1973spectral}
------, ``Spectral analysis of speech by linear prediction,'' \emph{IEEE Transactions on Audio and Electroacoustics}, vol.~21, no.~3, pp. 140--148, 1973.

\bibitem{lee1992robust}
K.~Y. Lee, B.-G. Lee, I.~Song, and S.~Ann, ``Robust estimation of ar parameters and its application for speech enhancement,'' in \emph{[Proceedings] ICASSP-92: 1992 IEEE International Conference on Acoustics, Speech, and Signal Processing}, vol.~1.\hskip 1em plus 0.5em minus 0.4em\relax IEEE, 1992, pp. 309--312.

\bibitem{rabiner1978digital}
L.~R. Rabiner, \emph{Digital processing of speech signals}.\hskip 1em plus 0.5em minus 0.4em\relax Pearson Education India, 1978.

\bibitem{fant1985four}
G.~Fant, J.~Liljencrants, Q.-g. Lin \emph{et~al.}, ``A four-parameter model of glottal flow,'' \emph{STL-QPSR}, vol.~4, no. 1985, pp. 1--13, 1985.

\bibitem{fujisaki1986proposal}
H.~Fujisaki and M.~Ljungqvist, ``Proposal and evaluation of models for the glottal source waveform,'' in \emph{ICASSP'86. IEEE International Conference on Acoustics, Speech, and Signal Processing}, vol.~11.\hskip 1em plus 0.5em minus 0.4em\relax IEEE, 1986, pp. 1605--1608.

\bibitem{klatt1990analysis}
D.~H. Klatt and L.~C. Klatt, ``Analysis, synthesis, and perception of voice quality variations among female and male talkers,'' \emph{the Journal of the Acoustical Society of America}, vol.~87, no.~2, pp. 820--857, 1990.

\bibitem{ding1995simultaneous}
W.~Ding, H.~Kasuya, and S.~Adachi, ``Simultaneous estimation of vocal tract and voice source parameters based on an arx model,'' \emph{IEICE transactions on information and systems}, vol.~78, no.~6, pp. 738--743, 1995.

\bibitem{fujisaki1996estimation}
H.~Fujisaki and M.~Ljungqvist, ``Estimation of voice source and vocal tract parameters based on arma analysis and a model for the glottal source waveform,'' in \emph{Recent Research Towards Advanced Man-machine Interface Through Spoken Language}.\hskip 1em plus 0.5em minus 0.4em\relax Elsevier, 1996, pp. 52--60.

\bibitem{vincent2005estimation}
D.~Vincent, O.~Rosec, and T.~Chonavel, ``Estimation of lf glottal source parameters based on an arx model,'' in \emph{Ninth European Conference on Speech Communication and Technology}, 2005.

\bibitem{li2020simultaneous}
Y.~Li, K.-I. Sakakibara, and M.~Akagi, ``Simultaneous estimation of glottal source waveforms and vocal tract shapes from speech signals based on arx-lf model,'' \emph{Journal of Signal Processing Systems}, vol.~92, no.~8, pp. 831--838, 2020.

\bibitem{li2021f_0}
Y.~Li, J.~Tao, D.~Erickson, B.~Liu, and M.~Akagi, ``$ f_0 $-noise-robust glottal source and vocal tract analysis based on arx-lf model,'' \emph{IEEE/ACM Transactions on Audio, Speech, and Language Processing}, vol.~29, pp. 3375--3383, 2021, doi: {10.1109/TASLP.2021.3120585}.

\bibitem{takahashi2018estimation}
K.~Takahashi and M.~Akagi, ``Estimation of glottal source waveforms and vocal tract shape for singing voices with wide frequency range,'' in \emph{2018 Asia-Pacific Signal and Information Processing Association Annual Summit and Conference (APSIPA ASC)}.\hskip 1em plus 0.5em minus 0.4em\relax IEEE, 2018, pp. 1879--1887.

\bibitem{rahman2007identification}
M.~S. Rahman and T.~Shimamura, ``Identification of arma speech models using an effective representation of voice source,'' \emph{IEICE transactions on information and systems}, vol.~90, no.~5, pp. 863--867, 2007.

\bibitem{ouaaline1998pole}
N.~Ouaaline and L.~Radouane, ``Pole--zero estimation of speech signal based on zero-tracking algorithm,'' \emph{International journal of adaptive control and signal processing}, vol.~12, no.~1, pp. 1--12, 1998.

\bibitem{morikawa1982adaptive}
H.~Morikawa and H.~Fujisaki, ``Adaptive analysis of speech based on a pole-zero representation,'' \emph{IEEE Transactions on Acoustics, Speech, and Signal Processing}, vol.~30, no.~1, pp. 77--88, 1982.

\bibitem{cadzow1980high}
J.~Cadzow, ``High performance spectral estimation--a new arma method,'' \emph{IEEE Transactions on Acoustics, Speech, and Signal Processing}, vol.~28, no.~5, pp. 524--529, 1980.

\bibitem{kopec1977speech}
G.~Kopec, A.~Oppenheim, and J.~Tribolet, ``Speech analysis homomorphic prediction,'' \emph{IEEE Transactions on Acoustics, Speech, and Signal Processing}, vol.~25, no.~1, pp. 40--49, 1977.

\bibitem{li2021study}
K.~Li, M.~Unoki, Y.~Li, J.~Dang, and M.~Akagi, ``Study on simultaneous estimation of glottal source and vocal tract parameters by armax-lf model for speech analysis/synthesis,'' in \emph{2021 Asia-Pacific Signal and Information Processing Association Annual Summit and Conference (APSIPA ASC)}.\hskip 1em plus 0.5em minus 0.4em\relax IEEE, 2021, pp. 36--43.

\bibitem{agiomyrgiannakis2009arx}
Y.~Agiomyrgiannakis and O.~Rosec, ``Arx-lf-based source-filter methods for voice modification and transformation,'' in \emph{2009 IEEE International Conference on Acoustics, Speech and Signal Processing}.\hskip 1em plus 0.5em minus 0.4em\relax IEEE, 2009, pp. 3589--3592.

\bibitem{fu2006robust}
Q.~Fu and P.~Murphy, ``Robust glottal source estimation based on joint source-filter model optimization,'' \emph{IEEE Transactions on Audio, Speech, and Language Processing}, vol.~14, no.~2, pp. 492--501, 2006.

\bibitem{alku1992glottal}
P.~Alku, ``Glottal wave analysis with pitch synchronous iterative adaptive inverse filtering,'' \emph{Speech communication}, vol.~11, no. 2-3, pp. 109--118, 1992.

\bibitem{drugman2012comparative}
T.~Drugman, B.~Bozkurt, and T.~Dutoit, ``A comparative study of glottal source estimation techniques,'' \emph{Computer Speech \& Language}, vol.~26, no.~1, pp. 20--34, 2012.

\bibitem{drugman2011detection}
T.~Drugman, M.~Thomas, J.~Gudnason, P.~Naylor, and T.~Dutoit, ``Detection of glottal closure instants from speech signals: A quantitative review,'' \emph{IEEE Transactions on Audio, Speech, and Language Processing}, vol.~20, no.~3, pp. 994--1006, 2011.

\bibitem{alku2019openglot}
P.~Alku, T.~Murtola, J.~Malinen, J.~Kuortti, B.~Story, M.~Airaksinen, M.~Salmi, E.~Vilkman, and A.~Geneid, ``Openglot--an open environment for the evaluation of glottal inverse filtering,'' \emph{Speech Communication}, vol. 107, pp. 38--47, 2019.

\bibitem{kawahara2016sparkng}
H.~Kawahara, ``Sparkng: Interactive matlab tools for introduction to speech production, perception and processing fundamentals and application of the aliasing-free lf model component.'' in \emph{INTERSPEECH}, 2016, pp. 1180--1181.

\bibitem{raitio2014voice}
T.~Raitio, H.~Lu, J.~Kane, A.~Suni, M.~Vainio, S.~King, and P.~Alku, ``Voice source modelling using deep neural networks for statistical parametric speech synthesis,'' in \emph{2014 22nd European Signal Processing Conference (EUSIPCO)}.\hskip 1em plus 0.5em minus 0.4em\relax IEEE, 2014, pp. 2290--2294.

\bibitem{kingma2014adam}
D.~P. Kingma and J.~Ba, ``Adam: A method for stochastic optimization,'' \emph{arXiv preprint arXiv:1412.6980}, 2014.

\bibitem{kane2013automating}
J.~Kane and C.~Gobl, ``Automating manual user strategies for precise voice source analysis,'' \emph{Speech Communication}, vol.~55, no.~3, pp. 397--414, 2013.

\bibitem{kane2012exploiting}
J.~Kane, I.~Yanushevskaya, A.~N{\'\i}~Chasaide, and C.~Gobl, ``Exploiting time and frequency domain measures for precise voice source parameterisation,'' in \emph{Speech Prosody 2012}, 2012.

\bibitem{becker2008forensic}
T.~Becker, M.~Jessen, and C.~Grigoras, ``Forensic speaker verification using formant features and gaussian mixture models,'' in \emph{Ninth Annual Conference of the International Speech Communication Association}, 2008.

\bibitem{tomashenko2022voiceprivacy}
N.~Tomashenko, X.~Wang, X.~Miao, H.~Nourtel, P.~Champion, M.~Todisco, E.~Vincent, N.~Evans, J.~Yamagishi, and J.~F. Bonastre, ``The voiceprivacy 2022 challenge evaluation plan,'' \emph{arXiv preprint arXiv:2203.12468}, 2022.

\bibitem{alku2011glottal}
P.~Alku, ``Glottal inverse filtering analysis of human voice production—a review of estimation and parameterization methods of the glottal excitation and their applications,'' \emph{Sadhana}, vol.~36, no.~5, pp. 623--650, 2011.

\bibitem{drugman2014glottal}
T.~Drugman, P.~Alku, A.~Alwan, and B.~Yegnanarayana, ``Glottal source processing: From analysis to applications,'' \emph{Computer Speech \& Language}, vol.~28, no.~5, pp. 1117--1138, 2014.

\end{thebibliography}

\end{document}